\documentclass[aps,preprint,nobibnotes,nofootinbib]{revtex4}
\usepackage{amsfonts}
\usepackage{graphicx}
\usepackage{amsmath}

\begin{document}

\title{Gauge and Averaging in Gravitational Self-force}

\author{Samuel E. Gralla}
\affiliation{\it Enrico
Fermi Institute and Department of Physics \\ University of Chicago
\\ \it 5640 S.~Ellis Avenue, Chicago, IL~60637, USA}

\begin{abstract}
A difficulty with previous treatments of the gravitational self-force is that an explicit formula for the force is available only in a particular gauge (Lorenz gauge), where the force in other gauges must be found through a transformation law once the Lorenz gauge force is known.  For a class of gauges satisfying a ``parity condition'' ensuring that the Hamiltonian center of mass of the particle is well-defined, I show that the gravitational self-force is always given by the angle-average of the bare gravitational force.  To derive this result I replace the computational strategy of previous work with a new approach, wherein the form of the force is first fixed up to a gauge-invariant piece by simple manipulations, and then that piece is determined by working in a gauge designed specifically to simplify the computation.  This offers significant computational savings over the Lorenz gauge, since the Hadamard expansion is avoided entirely and the metric perturbation takes a very simple form.  I also show that the rest mass of the particle does not evolve due to first-order self-force effects.  Finally, I consider the ``mode sum regularization'' scheme for computing the self-force in black hole background spacetimes, and use the angle-average form of the force to show that the same mode-by-mode subtraction may be performed in all parity-regular gauges.  It appears plausible that suitably modified versions of the Regge-Wheeler and radiation gauges (convenient to Schwarzschild and Kerr, respectively) are in this class.
\end{abstract}

\maketitle

\addtocounter{section}{1}

The leading-order deviation from geodesic motion proportional to the mass of a body---interpreted as the force due to the body's own gravitational field---is known as the gravitational self-force.  A recurring source of difficulty in both the theoretical treatment and practical computation of the self-force has been the choice of \textit{gauge} in which the metric perturbation and force are expressed.  In particular, early treatments \cite{mino-sasaki-tanaka,quinn-wald,detweiler-whiting,poisson-review} required a specific gauge choice---Lorenz gauge---even to define the perturbed trajectory (via a point particle hypothesis coupled with ``Lorenz gauge relaxation'' \cite{quinn-wald,gralla-wald} to allow non-geodesic motion), and a proposed extension of the results to other gauges \cite{barack-ori-gauge} restricted to gauge vectors that are continuous at the particle, even though the metric perturbation is singular.  At a theoretical level, this elevates a particular gauge to fundamental status, to the exclusion of other gauges that seem equally nice, such as any gauge where the point particle $1/r$ singularity corresponds to linearized Schwarzschild in Cartesian Schwarzschild coordinates, as opposed to the Cartesian isotropic coordinates that correspond to Lorenz gauge.\footnote{The (discontinuous) gauge vector that changes the singularity from isotropic to Schwarzschild type is given by $\xi^i=n^i=x^i/r$ in Fermi normal coordinates.}  And at a practical level, one has excluded the standard gauges of black hole perturbation theory \cite{barack-ori-gauge}.

Previous work \cite{gralla-wald} (hereafter paper I) eliminated the fundamental status of the Lorenz gauge by giving a definition of perturbed motion holding for any gauge where the particle is represented by a $1/r$ singularity.  However, in this work we still relied on the Lorenz gauge for our computations and, more importantly, for the \textit{expression} of the final result, as a formula holding in the Lorenz gauge together with a generalized transformation law.  At a theoretical level, we still have a preferred role for the Lorenz gauge; and at a practical level, the results suggest that the computation of a self-force in an alternative gauge must always proceed through Lorenz gauge, eliminating much of the appeal of working in alternative gauges in the first place.

In this paper I will identify a class of gauges based on the requirement that the center of mass as defined by Regge and Teitelboim \cite{regge-teitelboim} is well-defined (in the ``near zone''), and show that the force in any such gauge is given by the angle-average of the bare force in that gauge.  To derive this result I adopt the assumptions of paper I but take a new computational approach, wherein the form of the force in any gauge is fixed up to a gauge-invariant piece by simple manipulations, and then that piece is determined by working in a gauge chosen specifically to make the computation as simple as possible.  This approach avoids much of the computational complexity of previous work (eliminating the Hadamard expansion entirely and significantly reducing the calculation needed thereafter), while organizing the computation so that the final equation automatically takes a gauge-independent form.

The precise results are as follows.  We define the (lowest-order) mass $M$, spin $S_{ab}$, and center of mass deviation $Z^a$ of the particle as tensors on a timelike worldline $\gamma$ (four velocity $u^a$) in a vacuum background metric $g_{ab}$.\footnote{The spin is antisymmetric and satisfies $u^a S_{ab}=0$.  The deviation satisfies $Z^a u_a=0$.}  Then we find that that $\gamma$ is a geodesic, that the mass and spin are constant (parallel propagated), and that the deviation $Z^a$ satisfies
\begin{equation}\label{eom}
u^b \nabla_b (u^c \nabla_c Z^a) = \frac{1}{4\pi} \lim_{r\rightarrow 0}\int F^a d\Omega + R_{bcd}^{\  \ \ a} u^b Z^c u^d + \frac{1}{2M} R_{bcd}^{\ \ \ a} S^{bc} u^d,
\end{equation}
where $F^a$ is the ``bare gravitational force'', 
\begin{equation}\label{gravforce}
F^a = - \left(g^{ab}+u^a u^b\right)\left( \nabla_d h_{bc} - \frac{1}{2}\nabla_b h_{cd} \right)u^c u^d,
\end{equation}
and $h_{\mu \nu}$ is the metric perturbation of a point particle,
\begin{equation}\label{pp}
G^{(1)}_{ab}[h] = 8 \pi M \int_\gamma \delta_4 (x,z(\tau)) u_a u_b d\tau,
\end{equation}
which must be expressed in a ``parity-regular'' gauge, where the singular part of the spatial metric is even-parity on the sphere, $C_{ij}(t,-\vec{n})=C_{ij}(t,\vec{n})$ in equation \eqref{g1}.  The bare force $F^a$ is familiar from the perturbed geodesic equation, and here is defined only off of $\gamma$ (and only locally, where $u^a$ is extended off $\gamma$ by parallel transport along spacelike geodesics orthogonal to $u^a$) since the metric perturbation is divergent.  The integral in \eqref{eom} is defined by using the exponential map based on $\gamma$ to associate a flat metric, in terms of which the integral is over a fixed 2-sphere of spatial distance $r$ in the hyperplane orthogonal to $u^a$, with $n^a$ its unit normal and $d\Omega$ its area element, and the integration is done component-wise under the exponential map.  I also define the perturbed mass of the particle and show that it is constant in time.  

The first term on the right hand side of equation \eqref{eom} is proportional to the metric perturbation; I therefore refer to this term as the gravitational self-force.  We see that the force in any parity-regular gauge is given by the angle-average of the bare force in that gauge, so that the self-force may be viewed as the net gravitational force on the particle.  If the Lorenz gauge (which is parity-regular) is adopted and the Hadamard form for the metric perturbation is computed (choosing the retarded solution with no incoming radiation) and plugged in, then this term reduces to the standard ``tail integral'' expression for the self-force (e.g., \cite{poisson-review}).  The second term corresponds to the geodesic deviation equation and reflects the particle's desire to move on a new geodesic once it has been displaced from the original.  The third term is the Papapetrou spin force. If the parity condition is violated, equation \eqref{eom} does not hold, and the equation of motion takes a complicated form involving an explicit gauge transformation to a parity-regular gauge, equation \eqref{eom-noparity}.

A practical technique for computing self-forces in black hole background spacetimes is known as ``mode sum regularization'' (\cite{barack-ori-modesum} and many other references).  In this approach one numerically solves for the spherical harmonic modes of the metric perturbation (the sum over which diverges at the particle), and performs a mode-by-mode subtraction that regularizes the sum in such a way that the correct self-force is computed.  Extensive work has determined the form of the subtraction in the Lorenz gauge, which automatically holds for gauges that are smoothly related.  Taking advantage of a connection between mode decompositions and averaging (and hence self-force), I show that the same subtraction may in fact be performed in all parity-regular gauges.  It appears plausible that suitably modified Regge-Wheeler and radiation gauges (convenient to perturbations of Schwarzschild and Kerr, respectively) will be in this class, with especially strong evidence in the case of radiation gauge (see discussion at the end of section \ref{sec:eom}).  If so, then the results of this paper would provide a theoretical and practical foundation for the computation of self-force effects in gauges convenient to black hole spacetimes.

I use the conventions of Wald \cite{wald}.  Greek indices label tensor components, while early-alphabet Latin indices $a,b,\dots$ are abstract indices.  When coordinates $t,x^i$ are used, the time and space components are denoted by $0$ and mid-alphabet Latin indices $i,j,\dots$, respectively.

\section{Review of Formalism}\label{sec:review}

The central idea of paper I is to introduce a mathematically precise formulation of the notion of the ``near-zone'' of a body, and to use the requirement of a sensible near-zone to demand a sensible perturbation family.  Given a family of metrics $g_{ab}(\lambda)$ in coordinates ($t, x^i$), we define a scaled metric $\bar{g}_{ab} \equiv \lambda^{-2} g_{ab}$ and scaled coordinates $\bar{t} \equiv \lambda^{-1}(t-t_0)$ and $\bar{x}^i \equiv \lambda^{-1} x^i$.  Denoting the scaled metric in scaled coordinates by $\bar{g}_{\bar{\mu} \bar{\nu}}$, consider the $\lambda \rightarrow 0$ limit, $\bar{g}^{(0)}_{\bar{\mu} \bar{\nu}} \equiv \bar{g}_{\bar{\mu} \bar{\nu}}|_{\lambda=0}$.  This limit effectively ``zooms in'' on the spacetime point ($t = t_0, x^i = 0$), and will recover the near zone of a body if the one-parameter-family contains a body whose radius and mass shrink down linearly with $\lambda$ to the worldline $x^i = 0$ (denoted by $\gamma$). By demanding its existence (and associated conditions), we automatically consider bodies of small size and mass.  Note that the components of the original and scaled metrics are related simply by ``plugging in the new coordinates,''
\begin{equation}\label{plugin}
\bar{g}_{\bar{\mu} \bar{\nu}}(\lambda; t_0; \bar{t},\bar{x}^i)
= g_{\mu \nu} (\lambda; t=t_0 + \lambda \bar{t}, x^i=\lambda \bar{x}^i).
\end{equation}
This equation relates components of the scaled metric in scaled coordinates, $\bar{g}_{\bar{\mu} \bar{\nu}}$, to corresponding components of the original metric in the original coordinates, $g_{\mu \nu}$.

One may perform perturbation theory in either the original (``far-zone'') picture or the scaled (``near-zone'') picture.  The far-zone background and perturbations will be denoted by $g^{(n)}_{\mu \nu} \equiv (1/n!) (\partial_\lambda)^n g_{\mu \nu}|_{\lambda=0}$ (I also write $h_{\mu \nu}=g^{(1)}_{\mu \nu}$), while the near-zone background and perturbations will be denoted by $\bar{g}^{(n)}_{\bar{\mu} \bar{\nu}} \equiv (1/n!) (\partial_\lambda)^n \bar{g}_{\bar{\mu} \bar{\nu}}|_{\lambda=0}$.  Our assumptions together with the choice of Fermi coordinates (e.g., \cite{poisson-review}) about $\gamma$ for the far-zone background metric constrain the far-zone quantities to take the form
\begin{align}
g^{(0)}_{\mu \nu} & = \eta_{\mu \nu} + B_{\mu i \nu j}(t)x^i x^j + O(r^3) \label{g0} \\
h_{\mu \nu} = g^{(1)}_{\mu \nu} & = \frac{C_{\mu \nu}(t,\vec{n})}{r} + D_{\mu \nu}(t,\vec{n}) + r E_{\mu \nu}(t,\vec{n})+ O(r^2) \label{g1}
 \\
g^{(2)}_{\mu \nu} & = \frac{F_{\mu \nu}(t,\vec{n})}{r^2} + \frac{H_{\mu \nu}(t,\vec{n})}{r} + K_{\mu \nu}(t,\vec{n}) + O(r),\label{g2}
\end{align}
where we have defined $r=\sqrt{\delta_{ij}x^i x^j}$ and $n^i=x^i/r$.
Here $B_{\mu i \nu j}$ is related to the Riemann curvature on the background worldline $x^i=0$ (see e.g. \cite{poisson-review} for the exact expression), whereas $C_{\mu \nu},\dots,K_{\mu \nu}$ are unspecified smooth functions on $\mathbb{R} \times S^2$ (denoted by arguments $(t,\vec{n})$).  The lack of a linear term in equation \eqref{g0} is a consequence of $\gamma$ being geodesic, as shown in paper I.  We also showed that the metric perturbation has effective point particle source, equation \eqref{pp}.   In light of equation \eqref{plugin} the near-zone perturbation series takes the form
\begin{align}
\bar{g}^{(0)}_{\bar{\mu}\bar{\nu}} & = \eta_{\mu \nu} + \frac{C_{\mu \nu}(t_0,\vec{n})}{\bar{r}} + \frac{F_{\mu \nu}(t_0,\vec{n})}{\bar{r}^2} + O\left( \frac{1}{\bar{r}^3} \right) \label{gb0} \\
\bar{g}^{(1)}_{\bar{\mu} \bar{\nu}} & = D_{\mu \nu} + \frac{H_{\mu \nu}}{\bar{r}} + O\left( \frac{1}{\bar{r}^2} \right) + \bar{t} \left[ \frac{\dot{C}_{\mu \nu}}{\bar{r}} + \frac{\dot{F}_{\mu \nu}}{\bar{r}^2} + O \left( \frac{1}{\bar{r}^3} \right)
\right] \label{gb1} \\
\bar{g}^{(2)}_{\bar{\mu}\bar{\nu}} & = B_{\mu i \nu j}\bar{x}^i \bar{x}^j + \bar{r} E_{\mu \nu} + K_{\mu \nu} + O\left( \frac{1}{\bar{r}} \right) \nonumber \\ & + \bar{t} \left[ \dot{D}_{\mu \nu} + \frac{\dot{H}_{\mu \nu}}{\bar{r}} + O\left( \frac{1}{\bar{r}^2} \right) \right] + \frac{1}{2} \bar{t}^2 \left[ \frac{\ddot{C}_{\mu \nu}}{\bar{r}} + \frac{\ddot{F}_{\mu \nu}}{\bar{r}^2} + O\left( \frac{1}{\bar{r}^3} \right) \right], \label{gb2}
\end{align}
where an overdot denotes a $t$-derivative, and the order symbols $O(1/\bar{r}^n)$ refer to $\bar{t}$-independent functions.  The dependence on $(t_0,\vec{n})$ is suppressed in equations \eqref{gb1} and \eqref{gb2} for readability.   Note that the lack of growing-in-$\bar{r}$ terms in equation \eqref{gb1} is inherited from the absence of a linear term in \eqref{g0}, which is a consequence of $\gamma$ being geodesic (and choosing Fermi coordinates). The (stationary and asymptotically flat) near-zone background metric $\bar{g}^{(0)}_{\bar{\mu}\bar{\nu}}$ represents the local state of the body at time $t_0$, and as such its properties should characterize those of the body.  In particular, the multipole moments of $\bar{g}^{(0)}_{\bar{\mu}\bar{\nu}}$ should correspond to those of the body, at lowest non-trivial perturbative order.  To the (far-zone) perturbative order considered in this paper, only the monopole and dipole moments can play a role.  We therefore define\footnote{Equation \eqref{Sdef} for the spin holds only in coordinates where $C_{i0}$ vanishes.  A formula for the spin holding in general coordinates would take a more complicated form.}
\begin{align}
M(t_0) & \equiv \frac{-1}{8\pi} \lim_{\bar{r} \rightarrow \infty} \int n^i \partial_i \bar{g}^{(0)}_{00} \bar{r}^2 d\Omega \label{Mdef} \\
D^i(t_0) & \equiv \frac{3}{8\pi} \lim_{\bar{r} \rightarrow \infty} \int \bar{g}^{(0)}_{00} n^i \bar{r}^2 d\Omega \label{Ddef} \\
S_{ij}(t_0) & \equiv \frac{3}{8\pi} \lim_{\bar{r} \rightarrow \infty} \int \bar{g}^{(0)}_{0[i} n^{\ }_{j]} \bar{r}^2 d\Omega, \label{Sdef}
\end{align}
and refer to $M$, $D^i$, and $S_{ij}$ as the (lowest-order) mass, mass dipole, and spin (current dipole) of the body.\footnote{The mass dipole may also be thought of as the time-space component of the spin tensor, $S^{0i}$.  However, we will work with a spin tensor that is orthogonal to $\gamma$, $S^{0i}=0$, defining a separate mass dipole.}  In equations \eqref{Mdef}-\eqref{Sdef}, the bars on the indices of coordinate components of the near-zone metric have been dropped, and the integrals are taken on fixed coordinate two-spheres with respect to the ``flat'' volume element $d\Omega$.  It is well known (and easily checked) that the mass dipole may be set to zero by translating the coordinates, $\bar{x}^i \rightarrow \bar{x}^i - Z^i$, with 
\begin{equation}\label{Zdef}
Z^i(t_0) \equiv D^i(t_0)/M(t_0).
\end{equation}
This gives the new coordinates the interpretation of being mass centered, so that $Z^i$ represents the center of mass position of the body in the original coordinates.  Since the above translation corresponds to a first-order gauge transformation in the far zone (recall $\bar{x}^i=x^i/\lambda$), we identify $Z^i$ with the first-order deviation of the center of mass position from its background position $\gamma$.  While the mass, spin, and deviation are defined in the near zone ``at spatial infinity'' (as a function of the coordinate time $t_0$ along $\gamma$ at which the near-zone limit is taken), they may equally well be viewed as tensors defined in the far-zone ``at the spatial origin'', i.e., as tensor fields on $\gamma$.  It was shown in paper I that the mass and spin do not evolve with time $t_0$ (i.e., that they are parallel propagated tensor fields on $\gamma$), while a Lorenz-gauge equation of motion (together with a gauge transformation law) was derived for the deviation vector.

The analysis of this paper will require transformation properties of $Z^i$ not only under ordinary translations, but also under transformations of the form $\delta \bar{x}^i = \alpha^i(\vec{n}) + O(1/\bar{r})$, i.e., under angle-dependent translations, or supertranslations.  Using the well-known fact that $\bar{g}_{00}=-1+2M/\bar{r} + O(1/\bar{r}^2)$ for all vacuum solutions of the form \eqref{gb0}, we have by direct computation that $\delta \bar{g}_{00} = 2 M \alpha^i n_i / \bar{r}^2 + O(1/\bar{r}^3)$, so that $Z^i$ transforms as
\begin{equation}\label{changeZ}
\delta Z^i = \frac{3}{4\pi} \int \alpha^j n_j n^i d\Omega.
\end{equation}
Since the angle-average of a vector picks out an $\ell=1$, electric parity part, we may restrict consideration to $\alpha^i$ of the form $\alpha^i = B^j n_j n^i + C^i$ for constants $B^i$ and $C^i$.  The associated change in $Z^i$ is then simply $B^i+ C^i$.  Thus $Z^i$ changes under supertranslations with $B^i \neq 0$, in addition to its change under ordinary translations $\alpha^i=C^i$.  

\section{Coordinate Choices and Perturbed Mass}\label{sec:dM}

The mass, spin, and deviation of the body were defined at lowest-order in the near zone, so that perturbative corrections to these quantities would naturally be defined at first and higher orders in the near-zone.  However, while the background metric is stationary and asymptotically flat (so that its multipole moments are well-defined), the $n$'th-order perturbation may have growth in $n$ combined powers of $\bar{t}$ and $\bar{r}$ (c.f. paper I and equations (\ref{gb0}-\ref{gb1})).  It turns out, however, that at first order the situation is more under control.  As already noted, the choice of Fermi coordinates in the far-zone background---together with the fact that $\gamma$ is geodesic---eliminates growing-in-$\bar{r}$ terms from the first near-zone perturbation, so that the perturbation is asymptotically flat.  Furthermore, as I now show, the constancy of the mass and spin similarly allows one to eliminate all $\bar{t}/\bar{r}$ and $\bar{t}/\bar{r}^2$ terms, so that the perturbation is stationary to $O(1/\bar{r}^2)$.  To see this, recall that $\bar{g}^{(0)}_{\bar{\mu}\bar{\nu}}$ is a stationary and asymptotically flat metric, so that, introducing $\delta_{\mu \nu}=\textrm{diag}(1,1,1,1)$ and $t^\alpha = (1,0,0,0)$, it may (at each $t_0$) be put in the standard form \cite{MTW},
\begin{equation}\label{gb0form}
\bar{g}^{(0)}_{\bar{\mu} \bar{\nu}} = \eta_{\mu \nu} + \frac{2M}{\bar{r}}\delta_{\mu \nu} + \frac{M^2}{2 \bar{r}^2} (3 \eta_{\mu \nu}- t_\mu t_\nu) - \frac{4 n^i  t_{(\mu} S_{\nu)i}}{\bar{r}^2}  + O\left(\frac{1}{\bar{r}^3}\right),
\end{equation}
where $S_{i0}=0$ for all $t_0$ (implying $D^i=0$, so that the coordinates are mass centered and ``track'' the motion of the body), while a priori $M$ and $S_{ij}$ may depend on $t_0$, reflecting evolution of the mass and spin.  However, since the $M$ and $S_{ij}$ in \eqref{gb0form} do correspond to those defined in \eqref{Mdef} and \eqref{Sdef}, we know that these quantities are in fact independent of time $t_0$, as shown in paper I and remarked above.  Thus the near-zone background metric is identical (through order $O(1/\bar{r}^2)$) at all $t_0$, which implies by the form of \eqref{gb1} that the near-zone perturbation is independent of $\bar{t}$ through $O(1/\bar{r}^2)$, becoming simply
\begin{equation}\label{gb1form}
\bar{g}^{(1)}_{\bar{\mu} \bar{\nu}} =  D_{\mu \nu}(\vec{n}) + H_{\mu \nu}(\vec{n})\frac{1}{\bar{r}} + O\left(\frac{1}{\bar{r}^2}\right) + \bar{t} \ O\left(\frac{1}{\bar{r}^3}\right).
\end{equation}
Further simplification can now be made.  The perturbation $\bar{g}^{(1)}_{\bar{\mu}\bar{\nu}}$ must satisfy the (vacuum) linearized Einstein equation about the near-zone background $\bar{g}^{(0)}_{\bar{\mu}\bar{\nu}}$.  The quantity $D_{\mu \nu}$ appears at order $O(1/\bar{r}^2)$ (and higher) in the linearized Einstein tensor, while $H_{\mu \nu}$ appears first at $O(1/\bar{r}^3)$.  However, at these orders only the first two terms in the background \eqref{gb0form} will appear, so that the background is effectively Schwarzschild.  Therefore, when considering only the explicitly displayed terms in the perturbation \eqref{gb1form}, we may use well known results \cite{regge-wheeler,zerilli} for perturbations of the Schwarzschild spacetime.  In particular, since stationary perturbations scaling as $\bar{r}^0$ and $\bar{r}^{-1}$ (as in \eqref{gb1form}) are known to be $\ell=0$ (spherically symmetric) up to gauge, the perturbation in \eqref{gb1form} is simply a change in mass, and a gauge may be chosen so that
\begin{equation}\label{gb1form2}
\bar{g}^{(1)}_{\bar{\mu} \bar{\nu}} = \frac{2\ \delta M(t_0)}{\bar{r}}\delta_{\mu \nu} + O\left(\frac{1}{\bar{r}^2}\right) + \bar{t} \ O\left(\frac{1}{\bar{r}^3}\right),
\end{equation}
where $\delta M$ is an arbitrary constant (which here may a priori depend on $t_0$), which I refer to as the \textit{perturbed mass} of the body.  An explicit formula may be given in analogy with \eqref{Mdef},
\begin{equation}
\delta M(t_0) = \frac{-1}{8\pi} \lim_{\bar{r} \rightarrow \infty} \int n^i \partial_i \bar{g}^{(1)}_{00} \bar{r}^2 d\Omega \label{dMdef},
\end{equation}
with the caveat that this equation holds only in coordinates where $D_{\mu \nu}=\dot{C}_{\mu \nu}=0$.  If one wishes to compute the mass in other coordinates (such as when the Lorenz gauge is used, and $D_{\mu \nu}$ is equal to the value of the ``tail integral''), a more complicated expression (which must implicitly involve transforming to appropriate coordinates) must be derived.  However, since the mass is an intrinsic property of a spacetime, there is no need to consider such coordinates in defining $\delta M$ and determining its evolution.\footnote{A definition of perturbed mass was given in \cite{pound}, which appears to correspond to equation \eqref{dMdef} applied in the Lorenz gauge.  This definition would not be sensible within our framework.  The mass defined in \cite{pound} was found to evolve with time.  The conclusion that the perturbed mass evolves with time appears to be at odds with the analysis of \cite{quinn-wald-energy}, where it was found that energy conservation is satisfied under the assumption of no change in mass.}  Note that since the perturbation \eqref{gb1form} is still stationary at $O(1/\bar{r}^2)$, it may be possible to define perturbed spin and deviation by an analogous procedure.  However, these quantities appear at one order in $\lambda$ higher than pursued in this paper, and are not considered here.

I now show that the perturbed mass does not evolve with time.  With our previous coordinate choices, the second-order near-zone perturbation takes the from
\begin{equation}\label{gb2form}
\bar{g}^{(2)}_{\bar{\mu}\bar{\nu}} = B_{\mu i \nu j}\bar{x}^i \bar{x}^j + \bar{r} E_{\mu \nu} + K_{\mu \nu} + \frac{\bar{t}}{\bar{r}}2 \dot{\delta M} \delta_{\mu \nu} + O(1/\bar{r}) + \bar{t} \ O(1/\bar{r}^2) + \bar{t}^2 \ O(1/\bar{r}^3).
 \end{equation}
This perturbation satisfies the (vacuum) linearized Einstein equation about $\bar{g}_{\bar{\mu}\bar{\nu}}^{(0)}$ (equation \eqref{gb0form}) with effective sources constructed from $\bar{g}_{\bar{\mu}\bar{\nu}}^{(1)}$ (equation \eqref{gb1form2}).  However, it is easy to see that these effective sources are $O(1/\bar{r}^4)$, $\bar{t} O(1/\bar{r}^6)$, and $\bar{t}^2 O(1/\bar{r}^8)$, while the error in the linearized Einstein tensor introduced by including only the explicitly displayed terms in the second-order perturbation \eqref{gb2} is $O(1/\bar{r}^3)$, $\bar{t}O(1/\bar{r}^4)$, and $\bar{t}^2 O(1/\bar{r}^5)$.  Thus the effective source terms may be ignored, and denoting the linearized Einstein tensor of $\bar{g}^{(2)}_{\bar{\mu}\bar{\nu}}$ about $\bar{g}^{(0)}_{\bar{\mu}\bar{\nu}}$ by $\mathcal{G}_{\bar{\mu} \bar{\nu}}$, we have $\mathcal{G}_{\bar{\mu} \bar{\nu}}=0$ to the appropriate order, i.e., 
\begin{equation}
\mathcal{G}_{\bar{\mu} \bar{\nu}} = O(1/\bar{r}^3)+\bar{t}O(1/\bar{r}^4)+\bar{t}^2O(1/\bar{r}^5).
\end{equation}

To determine the mass evolution $\dot{\delta M}$ we focus on the $\ell=0$ part of $\mathcal{G}_{\bar{\mu} \bar{\nu}}$.  While the background $\bar{g}^{(0)}_{\bar{\mu}\bar{\nu}}$ is not spherically symmetric due to the presence of the spin term, it is easy to see that to the relevant order in $\mathcal{G}_{\bar{\mu} \bar{\nu}}$ the spin term only contributes in product with the ``B'' term in the perturbation \eqref{gb2form}.  However, since the far-zone background spacetime is assumed to be vacuum, its Riemann tensor may be decomposed into two rank-two symmetric, trace-free spatial tensors (the ``electric'' and ``magnetic'' parts, e.g. \cite{poisson-review}), showing that the ``B'' term is pure $\ell=2$.  Since the spin term is $\ell=1$, the combination can therefore make no contribution to the $\ell=0$ part of $\mathcal{G}_{\bar{\mu} \bar{\nu}}$ at the relevant order.  Instead, the $\ell=0$ part is completely determined by the remaining terms in the perturbation, which ``see'' only the Schwarzschild metric.  In particular only the $\ell=0$ parts of these terms may contribute to the $\ell=0$ part of $\mathcal{G}_{\bar{\mu} \bar{\nu}}$, and we conclude that the $\ell=0$ part of $\bar{g}^{(2)}_{\bar{\mu}\bar{\nu}}$ must be a perturbation of the Schwarzschild spacetime to the relevant order.  We may now use Zerilli's result \cite{zerilli} that $\ell=0$ perturbations of Schwarzschild simply shift the mass, being writable as a $1/\bar{r}$ term plus a gauge transformation.  Since $O(1/\bar{r})$ is higher than considered, we have that the $\ell=0$ part of $\bar{g}^{(2)}_{\bar{\mu}\bar{\nu}}$ is pure gauge to the relevant order,
\begin{equation}
\left( r E_{\mu \nu} + K_{\mu \nu} + \frac{\bar{t}}{\bar{r}} 2 \dot{\delta M} \delta_{\mu \nu} \right)_{\ell=0} = \stackrel{\textit{\tiny M}}{\nabla}_{(\mu} \xi_{\nu)} + O(1/\bar{r}) + \bar{t} \ O(1/\bar{r}^2) + \bar{t}^2 O(1/\bar{r}^3),
\end{equation}
where $\stackrel{\textit{\tiny M}}{\nabla}$ is the derivative operator compatible with Schwarzschild and $\xi^\mu$ is a vector field.  We may now take a time derivative to find
\begin{equation}\label{fancyG}
\partial_0 \stackrel{\textit{\tiny M}}{\nabla}_{(\mu} \xi_{\nu)} = \stackrel{\textit{\tiny M}}{\nabla}_{(\mu} \partial_0 \xi_{\nu)} = \frac{2 \dot{\delta M}}{\bar{r}} \delta_{\mu \nu} + O(1/\bar{r}^2) + \bar{t} \ O(1/\bar{r}^3),
\end{equation}
which immediately implies $\dot{\delta M}=0$, since the mass perturbation $2 \dot{\delta M}/\bar{r} \delta_{\mu \nu}$ is not pure gauge. 

Incorporating this result, we may now summarize the coordinate choices made in this section as  
\begin{align}
\bar{g}^{(0)}_{\bar{\mu} \bar{\nu}} & = \eta_{\mu \nu} + \frac{2M}{\bar{r}}\delta_{\mu \nu} + \frac{M^2}{2 \bar{r}^2} (3 \eta_{\mu \nu}- t_\mu t_\nu) - \frac{4 n^i  t_{(\mu} S_{\nu)i}}{\bar{r}^2}  + O\left(\frac{1}{\bar{r}^3}\right) \label{gb0gauged}\\
\bar{g}^{(1)}_{\bar{\mu} \bar{\nu}} & = \frac{2\delta M}{\bar{r}}\delta_{\mu \nu} + O\left(\frac{1}{\bar{r}^2}\right) + \bar{t} \ O\left(\frac{1}{\bar{r}^3}\right) \label{gb1gauged} \\
\bar{g}^{(2)}_{\bar{\mu}\bar{\nu}} & = B_{\mu i \nu j}\bar{x}^i \bar{x}^j + \bar{r} E_{\mu \nu} + K_{\mu \nu} + O\left(\frac{1}{\bar{r}}\right) + \bar{t} \ O\left(\frac{1}{\bar{r}^2}\right) + \bar{t}^2 \ O\left(\frac{1}{\bar{r}^3}\right),\label{gb2gauged}
\end{align}
with $S_{0i}=0$.  By fixing the mass and spin terms to a standard form at all $t_0$ and choosing the mass dipole to vanish for all $t_0$, the metric form has been made very simple, and all non-stationarity has been eliminated from the relevant orders in $\bar{r}$.  These properties make this gauge much simpler to use than the Lorenz gauge used in \cite{gralla-wald} and elsewhere.  Rewritten in the far-zone, the perturbation series in these coordinates becomes
\begin{align}
h_{\mu \nu} = g_{\mu \nu}^{(1)} & = \frac{2 M \delta_{\mu \nu}}{r} + E_{\mu \nu} r + O(r^2) \label{g1gauged} \\
g_{\mu \nu}^{(2)} & = \frac{M^2}{2 r^2} (3 \eta_{\mu \nu}- t_\mu t_\nu) - \frac{4 n^i  t_{(\mu} S_{\nu)i}}{r^2} + \frac{2\delta M}{r}\delta_{\mu \nu} + K_{\mu \nu} + O(r), \label{g2gauged}
\end{align}
with the background $g^{(0)}_{\mu \nu}$ still given by \eqref{g0}.

\section{Parity Condition}

The definition of center of mass adopted in section \ref{sec:review} is based on the dipole moment of the time-time component of a stationary, asysmptotically flat metric.  An alternative definition of center of mass is given by equation (5.13) of Regge and Teitelboim \cite{regge-teitelboim}, derived as the conserved quantity canonically conjugate to the asymptotic boost symmetry of asymptotically flat general relativity.\footnote{We note that a later formula due to Beig and O'Murchadha \cite{beig-omurchadha} is equivalent given the parity condition.}  Like the Hamiltonian notion of mass (normally referred to as the ADM mass), this ``Hamiltonian center of mass'' involves only the spatial metric, and is more general in that it may be applied to time-dependent spacetimes in addition to the stationary spacetimes we consider.  However, unlike the Hamiltonian notion of mass, the center of mass comes with an additional restriction: In order to ensure the existence of the integral defining the center of mass, Regge and Teitelboim impose a ``parity condition'' that the monopole $(1/r)$ part of the spatial metric be even parity, $C_{ij}(\vec{n})=C_{ij}(-\vec{n})$ in equation \eqref{gb0}.\footnote{Regge and Teitelboim also impose a parity condition on the extrinsic curvature.  However, this condition is not needed for the center of mass and plays no role in our analysis.}  This restriction is not necessary to define the mass dipole, which is finite for any metric of the form \eqref{gb0}.

Since the general metric form \eqref{gb0form} satisfies the parity condition, we see that the parity condition is simply a coordinate condition in the context of stationary, asymptotically flat spacetimes.  Rotations and translations will automatically preserve the parity condition, while a supertranslation $\delta x^i = \alpha^i(\vec{n}) + O(1/\bar{r})$ must now satisfy $\alpha^i = c^i + \Sigma^i(\vec{n})$, with $c^i$ constant and $\Sigma^i$ odd parity, $\Sigma^i(-\vec{n})=-\Sigma^i(\vec{n})$.  This form provides a natural split between the ``pure translation'' part $c^i$ and a ``pure supertranslation'' part $\Sigma^i$, which is odd-parity.  It is easily checked from equation \eqref{changeZ} that the mass dipole center of mass changes by $c^i$ under this transformation, so that the parity condition removes its supertranslation dependence.  In this case the transformation properties of the mass dipole center of mass agree with those established by Regge and Teitelboim for their center of mass; and since both notions give zero on the metric \eqref{gb0form}, the two definitions are equivalent in our context.  Therefore, the question of which definition to use is simply the question of whether to impose the parity condition.

One may take one of two alternative viewpoints on this matter.  First, since the formal Hamiltonian analysis yields a center of mass formula that diverges (in general) in coordinates that violate the parity condition, one may argue that such coordinates are ``too irregular'' to admit a notion of center of mass, even if the mass dipole formula is finite.  Alternatively, one may view the mass dipole formula as providing an extension of the Regge Teitelboim center of mass to a larger class of coordinates within the stationary case.  In any case, the parity condition adds a number of simplifying properties in the context of the present work: It eliminates the supertranslation dependence of the mass dipole, it allows the equations of motion to be expressed purely in terms of the local spacetime metric (see discussion in appendix \ref{sec:eom-noparity}), and it makes the mode-sum regularization scheme gauge invariant.

\section{Equation of Motion}\label{sec:eom}
In the gauge of section \ref{sec:dM}, the equation of motion for the deviation is simply $Z^i(t)=0$.  In principle, therefore, giving the change in $Z^i$ under a change in gauge provides the complete description of motion.  However, the more useful description of motion in other gauges, equation \eqref{eom}, may be derived as follows.  Begin with gauge transformations $x'^\mu = x^\mu + \lambda \xi^\mu$ of the form\footnote{In the appendix of paper I an opposite sign convention, $x'^\mu = x^\mu - \lambda \xi^\mu$, was used for the definition of the gauge vector.} 
\begin{equation}
\xi^\mu = \alpha^\mu(t,\vec{n}) + O(r), \label{xi} \\
\end{equation}
where $\xi^\mu$ is assumed smooth in $r$ at fixed $(t,\vec{n})$, so that in particular $\alpha^\mu$ is a smooth function of its arguments.  It is easy to check that such transformations preserve the form of equations (\ref{g0}-\ref{g1}), and thus are allowed under our assumptions.  As described in \cite{gralla-wald-gaugenote}, we furthermore believe (but have not proven) that expressibility in the form of equation \eqref{xi} is a necessary condition on an allowed transformation, except in the case of certain trivial one-parameter-families, where additional $\log r$ terms are allowed.  Thus we believe (but have not proven) that such transformations correspond precisely to the coordinate choices allowed by our formalism (not including the parity condition) at first order in $\lambda$, for non-trivial families of solutions.  In order to preserve the parity condition $C_{ij}(t,-\vec{n})=C_{ij}(t,\vec{n})$, we must restrict the form of $\alpha^\mu$ so that
\begin{equation}
\xi^i = c^i(t) + \Sigma^i(t,\vec{n}) + O(r), \label{xi-parity} \\
\end{equation}
with $\Sigma^i$ odd-parity, $\Sigma^i(t,-\vec{n})=-\Sigma^i(t,\vec{n})$.  I will refer to gauge transformations of the above form as parity-regular transformations, and I will define parity-regular gauges as those that are related to the gauge of section \ref{sec:dM} by a parity-regular transformation.  We believe that parity-regular gauges are the general class allowed by our assumptions plus the parity condition, except possibly in trivial cases.  Thus we believe that one may check if a given gauge is parity-regular by checking that the metric perturbation takes the form \eqref{g1} with $C_{ij}(t,-\vec{n})=C_{ij}(t,\vec{n})$.  However, absent a complete proof of the assertions in \cite{gralla-wald-gaugenote}, one must instead check that the gauge vector relating to a known-parity-regular gauge is of the form \eqref{xi-parity}.  At the end of this section I discuss the parity-regularity of some common gauge choices.

Under a change of gauge of the form \eqref{xi-parity}, the near-zone background coordinates change by $\bar{x}'^i=\bar{x}^i + c^i(t_0) + \Sigma^i(t_0,\vec{n}) + O(1/\bar{r})$.  Using equation \eqref{changeZ}, we see that the deviation $Z^i(t_0)$ changes by $c^i(t_0)$, as noted in the previous section.  We may express this in terms of the gauge vector $\xi^\mu$ by taking an angle-average over a small constant-$r$ sphere,\footnote{We could equivalently express $\delta Z^i$ as an average over a circle or over two antipodal points, since these averages all agree for a ``constant plus odd parity'' function.  The entire derivation of the equation of motion could then proceed unchanged, so that the self-force in fact may equivalently be written as the average of the bare force over a (constant geodesic distance) sphere, circle, or pair of points.}
\begin{equation}\label{deltaZ}
\delta Z^i = \langle \xi^i \rangle_{r \rightarrow 0} \equiv \frac{1}{4\pi} \lim_{r\rightarrow 0} \int \xi^i d\Omega.
\end{equation}
Equation \eqref{deltaZ} gives the change in deviation due to a parity-regular transformation made on any perturbation of the assumed form \eqref{g1}.

The key manipulation now follows.  Consider the second time derivative, $\delta \ddot{Z}^i = \partial_0 \partial_0 \delta Z^i$.  We have
\begin{align}
\delta \ddot{Z}_i & = \langle \partial_0 \partial_0 \xi_i \rangle_{r \rightarrow 0} \nonumber \\
& = \langle \nabla_0 \nabla_0 \xi_i + R_{0j0k}x^k \partial_j \xi_i + \nabla_0 \nabla_i \xi_0 - \nabla_i \nabla_0 \xi_0 + R_{i00}^{\ \ \ j}\xi_j \rangle_{r \rightarrow 0} \nonumber \\
& = \langle -( \nabla_0 \delta h_{0i} - \frac{1}{2} \nabla_i \delta h_{00}) \rangle_{r \rightarrow 0} - R_{0i0j} \langle{ \xi^j \rangle}_{r\rightarrow 0} \nonumber \\
& = \delta \langle F_i \rangle_{r \rightarrow 0} - R_{0i0j} \delta Z^j\label{dZdd}
\end{align}
In the second line we have rewritten in terms of covariant
derivatives ($R_{0k0l}x^l \partial_k \xi_i$ is a Christoffel term), as well as added zero
in the form of the Ricci identity.   However, since $\partial_j \xi_i$ is even parity to leading order and the Riemann tensor is smooth, the first Riemann term vanishes by virtue of the parity condition.  Noting that the remaining derivatives of $\xi^i$ appear only in symmetrized form, in the third line we reexpress in terms of the change in the metric perturbation, $\delta h_{\mu \nu} = -2 \nabla_{(\mu}\xi_{\nu)}$, finding precisely the gravitational force form of equation \eqref{gravforce}.  We also separate off the remaining Riemann term, which takes a geodesic deviation form.  In the last line we use equations \eqref{gravforce} and \eqref{deltaZ}, where the finiteness\footnote{The angle average of $F^i$ is manifestly finite in the gauge of section \ref{sec:dM}, and, by reversing the calculations of \eqref{dZdd}, may be easily seen to transform finitely (see also \eqref{Fchange}).} of $\langle F_i \rangle_{r \rightarrow 0}$ allows us to pull the $\delta$ out of the angle-average.  We may now ``drop the $\delta$'s'' to obtain
\begin{equation}\label{Zdd}
\ddot{Z}^i = \langle F^i \rangle_{r \rightarrow 0} - R_{0j0}^{\ \ \ i} Z^j + \mathcal{A}^i,
\end{equation}
where $\mathcal{A}^i$ is the constant of integration---an unknown
gauge-invariant acceleration.  Thus by simple manipulations we have fixed the form of the equation of motion up to a gauge-invariant piece, and may now work in any convenient (parity-regular) gauge to determine $\mathcal{A}^i$.

Since the gauge of section \ref{sec:dM} has $Z^i(t)=0$ it immediately eliminates two terms in \eqref{Zdd}, giving simply
\begin{equation}\label{Finsimplegauge}
\mathcal{A}^i = - \langle F^i \rangle_{r \rightarrow 0}
\end{equation}
in this gauge.  The interpretation of the gauge is that the gravitational self-force is exactly opposite to the gauge-invariant force, so that the total force is zero, and there is no deviation from geodesic motion ($Z^i(t)=0$).  Since the angle-average of a three-vector picks out an $\ell=1$, electric parity part, we need consider only the $\ell=1$, electric parity part of the bare force $F^i$ in order to compute $\mathcal{A}^i$ from equation \eqref{Finsimplegauge}.  Since only the ``E'' term in the perturbation \eqref{g1gauged} contributes, we may focus on the $\ell=1$, electric parity part of $E_{\mu \nu}$.  To do so, we return to second-order near-zone perturbation theory.

As remarked above in the derivation of the constancy of the perturbed mass, the relevant terms in the second-order metric perturbation $\bar{g}^{(2)}_{\bar{\mu}\bar{\nu}}$ satisfy the linearized Einstein equation off of the $\bar{g}^{(0)}_{\bar{\mu}\bar{\nu}}$ to the relevant order, equation \eqref{fancyG}.  As further remarked, the $\ell=1$ spin term in the background appears only in combination with the $\ell=2$ ``B'' term in the perturbation.  While there is no contribution to the $\ell=0$ mode relevant for the perturbed mass, there can be a contribution to the $\ell=1$ mode relevant for the deviation, which significantly complicates the analysis (see appendix \ref{sec:spin}).  However, if the spin is assumed to be zero from the outset, then, as in the mass evolution case, the metric perturbation ``sees'' only Schwarzschild to the relevant order, and we can make use of Zerilli's \cite{zerilli} analysis of perturbations of Schwarzschild.  In particular, Zerilli showed that $\ell=1$, electric parity perturbations are pure gauge, so that by a (second-order near-zone) gauge transformation we may eliminate the $\ell=1$, electric parity part of $E_{\mu \nu}$ entirely, whence it immediately follows from \eqref{Finsimplegauge} that $\mathcal{A}^i=0$.  Thus the equation of motion in the spinless case may be derived with very little effort, involving only the few lines of algebra of equation \eqref{dZdd}.  If the spin is not assumed zero, more algebra is required (though still significantly less than needed when the Lorenz gauge is used).  This case is treated in appendix \ref{sec:spin}, leading to   
\begin{equation}\label{AequalsS}
M \mathcal{A}_i = \frac{1}{2}S^{kl}R_{kl0i},
\end{equation}
showing that $\mathcal{A}^i$ is simply the acceleration due to the Papapetrou spin force.  We have thus justified the final equation of motion, which appears in covariant form in \eqref{eom}.

I now discuss the parity-regularity of some common gauge choices.  The Lorenz gauge is convenient for local series expansions about the particle.  From equation \eqref{g1gauged}, the gauge of section \ref{sec:dM} already satisfies the Lorenz condition at leading and subleading order, $\nabla^\mu (h_{\mu \nu} -(1/2) h g_{\mu \nu}) = O(1)$.   It is then easy to check that the gauge vector to a full Lorenz gauge must take the form $\xi^i = C^i(t) + O(r)$, so that the Lorenz gauge is parity-regular. 

The Regge-Wheeler gauge is convenient for perturbations of Schwarzschild.   Barack and Ori \cite{barack-ori-gauge} show that in a few specific cases the gauge vector takes the form \eqref{xi} (bounded but direction dependent), and their formulae also imply that in these cases one can choose the vector to satisfy the required parity property, equation  \eqref{xi-parity}.  However, Hopper and Evans \cite{hopper-evans} have shown that in general the Regge-Wheeler gauge metric perturbations contain a delta-function in the Schwarzschild radial coordinate at the position of the particle, so that in general the Regge-Wheeler gauge is too singular to define the motion by our procedure.  However, using the explicit formulae for the coefficient of the delta function found in \cite{hopper-evans}, it should be possible to simply eliminate the delta function by a gauge transformation during the process of reconstructing the metric perturbations.  If the resulting gauge is parity-regular (a suggestion consistent with Barack and Ori's restricted results), then the results of this paper would enable self-force computations to be made within the Regge-Wheeler formalism, using mode sum regularization if desired.

The radiation gauge is convenient for perturbations of Kerr.  In \cite{k} it was checked that a radiation gauge may be chosen so that the metric perturbations near the particle take the form $(1/r) C_{\mu \nu}(t,\vec{n})$, with $C_{ij}(t,\vec{n})$ even parity on the sphere, so that the singularity is properly $1/r$ and the parity condition is satisfied.\footnote{At the time of the writing of \cite{k}, I believed (and communicated to the authors of \cite{k}) that only this check was required for the results of the present paper to hold.}  However, it was not checked that $C_{\mu \nu}(t,\vec{n})$ is \textit{smooth} (as assumed in this paper), and in fact it can be seen from the analysis of \cite{k2} that $C_{\mu \nu}(t,\vec{n})$ contains a jump discontinuity along a great circle, which is inherited from a discontinuity in the metric perturbation located at the radial coordinate of the particle.  Such a discontinuity seems unlikely to threaten the validity of the results, since all of the relevant formulae remain defined.  However, in order to be certain that the results of this paper may be applied to the gauge of \cite{k}, a careful analysis of the gauge vector relating to some parity-regular gauge must be performed.  Armed with the explicit form of such a vector, it should be straightforward to check if the arguments of this paper still hold.

\section{Mode-Sum regularization}\label{sec:modesum}

The computation of gravitational self-forces on black hole background spacetimes is an important problem for gravitational-wave astronomy of extreme mass-ratio inspirals (e.g., \cite{barack-review}).  The angle-average formula suggests a straightforward way of proceeding: first numerically compute the metric perturbations of a point particle in any parity-regular gauge, and then perform an average to determine the force in that gauge.  However, while simple in principle, such a procedure may be difficult to carry out in practice, due to the singular nature of the quantity being averaged.  Instead, to achieve an accurate result it is likely preferable to use an alternative technique, such as that of \textit{mode sum regularization}, first introduced in \cite{barack-ori-modesum} and widely employed thereafter.  This method takes advantage of the fact that numerical calculations in black hole spacetimes usually employ a spherical (or spheroidal) harmonic decomposition, which in particular has the property that the individual modes of the bare force are finite at the particle.  One then performs a finite subtraction on each mode, which is designed so that the resulting sum converges to the correct self-force.  Extensive work has determined the form of this subtraction (in terms of ``regularization parameters'') for arbitrary orbits of Schwarzschild and Kerr in the Lorenz gauge.  In this section I show that the mode sum regularization scheme is gauge-invariant under the parity condition, in the sense that the same subtraction may be employed to determine the force in any parity-regular gauge.  Combined with the Lorenz gauge results of \cite{barack-ori-kerr}, this provides a complete regularization prescription for Kerr in parity-regular gauges.

Let $(\tilde{t},\tilde{r},\hat{\theta},\hat{\phi})$ be Boyer-Lindquist coordinates for the Kerr spacetime.  For a given point along the worldline $\gamma$ where we wish to compute the self-force, we may rotate and time-translate the coordinates so that the particle is located at $\tilde{t}=\tilde{\phi}=0$, while taking the remaining coordinate positions to be $\tilde{r}=r_0$ and $\hat{\theta}=\theta_0$.  Despite the lack of a full rotational symmetry we (following \cite{barack-ori-modesum}) nevertheless also perform an additional rotation in the $\hat{\theta}$ direction, so that the particle is located at the pole of the new coordinates.   More precisely, define new angular coordinates $(\tilde{\theta},\tilde{\phi})$ by
\begin{align}
\cos \hat{\theta} & = - \sin \tilde{\theta} \cos \tilde{\phi} \cos \theta_0 + \cos \tilde{\theta} \cos \theta_0 \label{theta-hat} \\ 
\tan \hat{\phi} & = \frac{\sin \tilde{\theta} \sin \tilde{\phi}}{\sin \tilde{\theta} \cos \tilde{\phi} \cos \theta_0 + \cos \tilde{\theta} \sin \theta_0} \label{phi-hat}
\end{align}
to obtain ``rotated Boyer-Lindquist'' coordinates $\tilde{t},\tilde{r},\tilde{\theta},\tilde{\phi}$ in which the particle position is given by $\tilde{t}=0,\tilde{r}=r_0,\tilde{\theta}=0$.  In these coordinates the metric components are smooth everywhere except for the pole $\theta=0$, where they acquire non-trivial direction-dependent limits.  Below we will need the lowest-order  relationship between the spatial Fermi coordinates $x^i$ and the rotated Boyer-Lindquist coordinates, restricted to the sphere.  A straightforward computation gives this to be
\begin{equation}\label{fermi-tilde}
x^i|_{\tilde{t}=0, \tilde{r}=r_0} = \alpha^i \ \tilde{\theta} \cos \tilde{\phi} + \beta^i \ \tilde{\theta} \sin \tilde{\phi} + O(\tilde{\theta}^2),
\end{equation} 
where $\alpha^i$ and $\beta^i$ are constants independent of $\tilde{\theta}$ and $\tilde{\phi}$ (dependent on the Boyer-Lindquist position $\theta_0$, the three-velocity, and the mass and spin parameters of the Kerr metric), and where the $O(\tilde{\theta}^2)$ term may depend on $\tilde{\phi}$.

The advantage of placing the particle at the pole is the simplification of the spherical harmonic description by the elimination of modes with non-zero $m$ when the series is evaluated at the particle.  Let a subscript $\ell$ denote the $\ell$'th term in the expansion evaluated at the pole/particle, $A_{\ell} = \int A Y_{\ell 0} d\tilde{\Omega}$ for integrable functions $A(\tilde{\theta},\tilde{\phi})$, and suitably generalized for distributions.  When viewed in light of our angle-average result, the mode sum regularization prescription amounts to finding some $S_\ell^i$ such that
\begin{equation}\label{modesum}
\langle F^i \rangle_{r \rightarrow 0} = \sum_{\ell=0}^\infty \left( F^i_{\ell} - S^i_{\ell} \right).
\end{equation}
\begin{figure}[t]
\includegraphics[scale=.5]{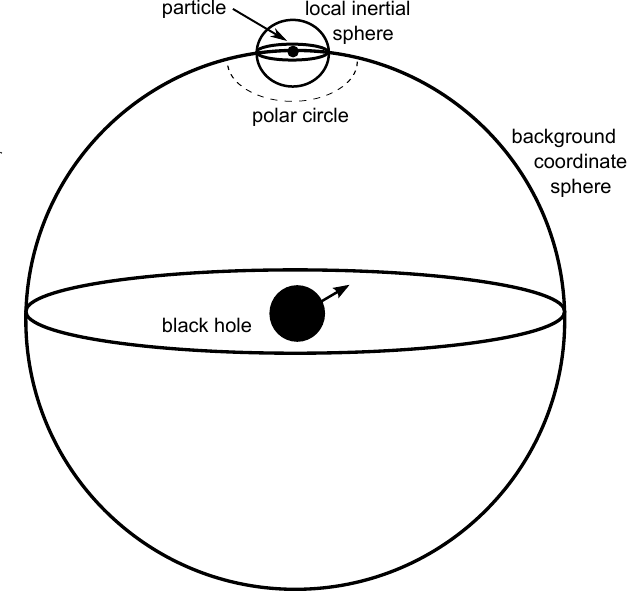}
\caption{A diagram illustrating the geometrical setup of the mode sum regularization argument.  The particle is at the pole of rotated Boyer-Lindquist coordinates.  The mode decomposition is taken relative to the background coordinate sphere, while the self-force is given by an average over the local inertial sphere.  For the change in bare force, a general theorem relates the mode sum at the particle/pole to the average over the polar circle, which agrees with the average over the local inertial sphere when the parity condition is satisfied.}
\label{fig:setup}
\end{figure}
This formula relates an average over an infinitesimal sphere surrounding the particle to a spherical harmonic decomposition on a finite sphere surrounding the black hole, evaluated at the particle (see figure \ref{fig:setup}).  Such a connection between mode sums and local averaging is familiar from ordinary Fourier series, where, if a function is of bounded variation, its series converges to the two-sided average at a discontinuity.  For spherical harmonic expansions, an analogous result (e.g. section III22b of \cite{sansone}) states that if the average of a function over latitude lines is of bounded variation (as a function of latitude), then its spherical harmonic series evaluated at the pole (``Laplace series'') converges to the average on an infinitesimal latitude line surrounding the pole.  (This result is easily understood at a formal level, by noting that $Y_{\ell 0}$ is independent of $\tilde{\phi}$, so that the formula for $A_{\ell}$ takes an average over $\tilde{\phi}$.)  The theorem does not apply to the bare force $F^i$ (which is divergent), but it does apply to the \textit{change} in bare force under a change in gauge, $\delta F^i$.  In particular, a simple calculation (which reverses the calculations internal to the average in \eqref{dZdd}) gives
\begin{equation}
\delta F^i = \partial_0 \partial_0 \xi^i + R^{\ i}_{0 \ 0j}\xi^j - R^{\ k}_{0 \ 0l} x^l \partial_k \xi^i + O(r) \label{Fchange}
\end{equation}
for any transformation of the form \eqref{xi}.  If the transformation is parity-regular, equation \eqref{xi-parity}, we see that $\delta F^i$ has a Fermi coordinate expansion of the form
\begin{equation}\label{Fchange-parity}
\delta F^i = \mathcal{C}^i(t) + \mathcal{S}^i(t,\vec{n}) + O(r),
\end{equation}
where $\mathcal{S}^i$ is smooth and odd-parity, $\mathcal{S}^i(t,-\vec{n})=-\mathcal{S}^i(t,\vec{n})$.\footnote{The parity condition is not required to show the applicability of the theorem, but will be necessary for the later analysis of this section.}  Restricting to the background coordinate sphere $\tilde{t}=0,\tilde{r}=r_0$ and expanding in $\tilde{\theta}$ at fixed $\tilde{\phi}$, we have
\begin{align}
\delta F^i|_{\tilde{t}=0,\tilde{r}=\tilde{r}_0}(\tilde{\theta},\tilde{\phi}) & = \mathcal{C}^i(0) + \mathcal{S}^i(0,\vec{n}|_{\tilde{t}=0,\tilde{r}=\tilde{r}_0}(\tilde{\theta},\tilde{\phi})) +  O(\tilde{\theta}) \label{starthere}\\
& = \mathcal{C}^i(0) + \mathcal{S}^i(0,\vec{n}|_{\tilde{t}=0,\tilde{r}=\tilde{r}_0}(\tilde{\theta}=0,\tilde{\phi})) +  O(\tilde{\theta}) \label{gethere}\\
& = \mathcal{C}^i(0) + \mathcal{S}^i(0,\vec{n}_0(\tilde{\phi})) + O(\tilde{\theta}),\label{deltaFexp}
\end{align}
where the $O(\tilde{\theta})$ terms may depend on $\tilde{\phi}$ and we have defined
\begin{equation}\label{n0}
\vec{n}_0(\tilde{\phi}) = \lim_{\tilde{\theta}\rightarrow 0} \vec{n}(\tilde{t}=0,\tilde{r}=\tilde{r}_0,\tilde{\theta},\tilde{\phi}).
\end{equation}
In moving from equation \eqref{starthere} to equation \eqref{gethere} we have used the fact that the restriction of $\vec{n}$ to the background coordinate sphere is smooth in $\tilde{\theta}$ at fixed $\tilde{\phi}$ (see equation \eqref{fermi-tilde} and recall $n^i=x^i/r$), as well as the fact that $\mathcal{S}^i$ is smooth in $\vec{n}$.  Equation \eqref{deltaFexp} shows that the restriction of $\delta F^i$ to the background coordinate sphere is continuous (in fact smooth, by our assumptions) in $\tilde{\theta}$ at fixed $\tilde{\phi}$.  In particular its average over $\tilde{\phi}$ is of bounded variation, and by the theorem we have
\begin{align}\label{poleavg}
\sum_{\ell} (\delta F^i)_\ell & = \lim_{\tilde{\theta}\rightarrow 0} \frac{1}{2 \pi} \int \delta F^i|_{\tilde{t}=0,\tilde{r}=r_0}(\tilde{\theta},\tilde{\phi}) d\tilde{\phi} \\
& = \mathcal{C}^i(0) + \frac{1}{2 \pi} \int \mathcal{S}^i(0,\vec{n}_0(\tilde{\phi})) d\tilde{\phi},
\end{align}
where in the second line we have plugged in the form of equation \eqref{deltaFexp}.  Since $\mathcal{S}^i$ is odd parity, the second term on the right-hand-side will vanish if $\vec{n}_0$ is odd under $\tilde{\phi} \rightarrow \tilde{\phi} + \pi$.  This property is expected from the geometry of the setup (figure \ref{fig:setup}), and is easily confirmed from equations \eqref{n0} and \eqref{fermi-tilde}. Thus the term involving $\mathcal{S}^i$ vanishes, so that the Laplace series for $\delta F^i$ in fact converges to $\mathcal{C}^i$.  However, the angle-average that computes the self-force also returns $\mathcal{C}^i$ on the form \eqref{Fchange-parity}.  Therefore, when the parity condition is satisfied the averages agree, and we have simply 
\begin{equation}\label{sumavg}
\sum_\ell (\delta F^i)_\ell = \langle \delta F^i \rangle_{r \rightarrow 0},
\end{equation}
showing that the Laplace series for the change in bare force $\delta F^i$ in fact converges to its local inertial angle-average, i.e., to the change in self-force it effects.  This means in particular that no extra $S^i_\ell$ must be subtracted in the new gauge, since merely the process of decomposing $\delta F^i$ into modes and resumming returns its contribution to the self-force.  To see this explicitly, let $F^i_\textrm{old}$ denote the bare force in a gauge that satisfies the parity condition, and write 
\begin{align}
\sum_\ell\left( F^i_\ell - S^i_\ell \right) & = \sum_\ell\left[ (F^i_{\textrm{old}})_\ell + (\delta F^i)_\ell - S_{\ell}^i \right] \nonumber \\
& = \sum_\ell\left[ (F^i_{\textrm{old}})_\ell - S_{\ell}^i \right] + \langle \delta F^i \rangle_{r \rightarrow 0} \nonumber \\
& = \langle F^i_{\textrm{old}}+\delta F^i \rangle_{r \rightarrow 0}. \label{showedit} 
\end{align}
In writing the second line we have used \eqref{sumavg}, and in writing the third line we have used equation \eqref{modesum}.  This shows that equation \eqref{modesum} holds in the new gauge if it held in the old, i.e., that $S^i_\ell$ is a correct piece to subtract in any parity-regular gauge.

We note that previous work has organized the subtraction so that one first subtracts an $\hat{S}^i_\ell$ of the form $\hat{S}^i_\ell = A^i (\ell + 1/2) + B^i + C^i/(\ell + 1/2)$, then sums over modes (the result is now finite), and finally adds in a finite residual $D^i$ to get the correct self-force.  (In terms of our $S^i_\ell$, $D^i$ is a ``finite piece'' $D^i \equiv \sum_\ell (S_\ell^i - \hat{S}_\ell^i)$.\footnote{Remarkably, it has been found (by lengthy computation in the Lorenz gauge) that $D^i=0$ in every circumstance, so that the subtraction of $A^i (\ell + 1/2) + B^i + C^i/(\ell + 1/2)$ in fact returns the correct force.  This surprising relationship between the large-$\ell$ expansion of a point particle metric perturbation (which uniquely determines $A,B,C$) and the physical self-force has thus far defied a more fundamental explanation.})  The ($\ell$-independent) $A,B,C,D$ are the ``regularization parameters'' for the particular orbit and spacetime, which by the results of this section do not depend on the choice of parity-regular gauge.

\begin{acknowledgements}
I gratefully acknowledge Robert Wald and John Friedman for helpful conversations, Laszlo Szabados for helpful correspondence, and an anonymous referee for a helpful comment.  This research was supported in part by NSF grant PHY08-54807 to the University of Chicago.
\end{acknowledgements}

\appendix

\section{Derivation in the case of non-zero spin}\label{sec:spin}

When the spin is non-zero, we may not rely on Zerilli's results for Schwarzschild \cite{zerilli}, and the analysis of second-order Einstein equation, equation \eqref{fancyG}, becomes more complicated.  In this case it pays to systematically consider the contributions to the linearized Einstein tensor $\mathcal{G}_{\bar{\mu}\bar{\nu}}$ from the various terms in the background \eqref{gb0gauged} and perturbation \eqref{gb2gauged}.  Since all terms are stationary to the relevant orders, no $\bar{t}$-dependence will appear, and we may count orders in $1/\bar{r}$.  At leading order $O(1)$ in $\mathcal{G}_{\bar{\mu}\bar{\nu}}$, the only contributions are from the ``B'' term in the perturbation \eqref{gb2gauged} and the flat ``$\eta$'' term in the background.  Since the ``B'' term is (by the Fermi coordinate construction) a perturbation of flat spacetime, the linearized Einstein equation is automatically satisfied and we learn no new information.  At next order $O(1/\bar{r})$ in $\mathcal{G}_{\bar{\mu}\bar{\nu}}$, both the ``E'' term in the perturbation and the mass term $2M \delta_{\mu \nu}/\bar{r}$ in the background can now contribute.  Expanding the background metric in powers of $1/\bar{r}$, the linearized Einstein equation at $O(1/\bar{r})$ may be written
\begin{equation}\label{moo}
G^{(1)}_{\eta}[E_{\mu \nu} \bar{r}] + 2 G^{(2)}_{\eta}[2M/\bar{r}\delta_{\mu \nu},B_{\mu i \nu j}\bar{x}^i \bar{x}^j] = 0,
\end{equation}
where $G^{(1)}_{\eta}$ and $G^{(2)}_{\eta}$ are the first and second-order Einstein tensors (respectively) off of flat spacetime.  However, since the ``B'' term has no $\ell=1$, electric parity part while the $M$ term is spherically symmetric, the second term in \eqref{moo} has no $\ell=1$, electric parity part.  Therefore the $\ell=1$, electric parity part of the ``E'' term satisfies the vacuum linearized Einstein equation about flat spacetime,
\begin{equation}\label{oink}
G^{(1)}_{\eta}[(E_{\mu \nu} \bar{r})_{\ell=1,+}] = 0,
\end{equation}
where the subscript ``$\ell=1,+$'' indicates the $\ell=1$, electric parity part.  Since $\ell=1$, electric parity perturbations of flat spacetime that scale linearly with $\bar{r}$ are pure gauge (e.g., \cite{zhang}), all solutions to equation \eqref{oink} may be written $(E_{\mu \nu} \bar{r})_{\ell=1,+}=\partial_{(\mu} \mathcal{E}_{\nu)}$, where $\mathcal{E}_\mu$ is  $\ell=1$ and electric parity.  If we make a (second-order near-zone) gauge transformation generated by $\mathcal{E}$, then we may set the $\ell=1$, electric parity part of $E_{\mu \nu}$ to zero.  However, one can check that in general the required gauge vector $\mathcal{E}_\mu$ is $\bar{t}$-dependent (despite the ``E'' term being stationary), so that this gauge transformation would introduce $\bar{t}$-dependence at higher orders, in violation of our previous choices that eliminated such dependence.  (In particular, terms of order $\bar{t}/\bar{r}$ and $\bar{t}^2/\bar{r}^2$ would appear in $\bar{g}^{(2)}_{\bar{\mu}\bar{\nu}}$, contradicting the previous choices $\dot{H}_{\mu \nu}=\ddot{C}_{\mu \nu}=0$---see equations \eqref{gb2} and \eqref{gb2gauged}.) Without invalidating these choices we may only make gauge transformations whose gauge vector is $\bar{t}$-independent.  One may check that this allows us to put $E_{\mu \nu}$ in the form
\begin{equation}\label{cluck}
(E_{\mu \nu} \bar{r})_{\ell=1,+} = -2 a_i \bar{x}^i t_\mu t_\nu,
\end{equation}
where $a_i$ is an arbitrary spatial vector, named since in this form $(E_{\mu \nu} \bar{r})_{\ell=1,+}$ is an ``acceleration perturbation'' familiar from Fermi coordinates about an accelerated worldline.  Making such a gauge transformation (and ``absorbing'' its effects at $O(1)$ into the arbitrary tensor $K_{\mu \nu}$), equation \eqref{Finsimplegauge} becomes simply
\begin{equation}\label{Finsimplegauge2}
\mathcal{A}^i = a^i.
\end{equation}

We have now made coordinate choices that eliminate all $\bar{t}$-dependence and reduce the relevant $\ell=1$, electric parity part of $E_{\mu \nu}$ into a simple form with one unknown, the gauge-invariant acceleration $\mathcal{A}^i=a^i$.  It remains to use the linearized Einstein equation at order $O(1/\bar{r}^2)$ to determine $\mathcal{A}^i$.  Again expanding the background in powers of $1/\bar{r}$, at this order the $\ell=1$, electric parity parts of the second-order Einstein equation may be written as
\begin{equation}
\left( G^{(1)}_{\eta}\left[K_{\mu \nu}\right] + 2 G^{(2)}_{\eta}\left[\frac{2M \delta_{\mu \nu}}{\bar{r}}, -2\mathcal{A}_i \bar{x}^i t_{\mu} t_{\nu}\right] + 2 G^{(2)}_{\eta}\left[\frac{-4 n^i t_{(\mu} S_{\nu)i}}{\bar{r}^2},B_{\mu i \nu j}\bar{x}^i \bar{x}^j\right] \right)_{\ell=1,+}=0,
\end{equation}
where terms that can give no $\ell=1$, electric part have not been displayed.  This equation gives relationships between $K_{\mu \nu}$, $\mathcal{A}_i$, $M$, $S_{ij}$, and $R_{\mu i \nu j}$.  To determine a relationship not involving the unknown tensor $K_{\mu \nu}$, first write out the linearized Einstein tensor about flat spacetime for stationary perturbations, and note that there is a particular $\ell=1$ part\footnote{In the notation of \cite{gralla-wald}, this part is $(2/3) R^D_i + R^E_i$, involving only spatial trace-free components of the linearized Ricci tensor.} that vanishes for all $K_{\mu \nu}$.  Then computing this particular $\ell=1$ part of the remaining two terms will determine $\mathcal{A}_i$ in terms of the mass, spin, and Riemann tensor.  Performing this calculation yields $M \mathcal{A}_i = \frac{1}{2}S^{kl}R_{kl0i}$, as claimed in equation \eqref{AequalsS}.
\section{Equation of motion in parity-irregular gauges}\label{sec:eom-noparity}

I now consider the form of the equations of motion in parity-irregular gauges, adopting the mass dipole definition of center of mass.  Under the general gauge transformations \eqref{xi}, we have from equation \eqref{changeZ} that\footnote{As already noted in section \ref{sec:eom}, the sign convention for $\xi^\mu$ used in this paper differs from that used in the appendix of paper I.}
\begin{equation}
\delta Z^i = \frac{3}{4\pi} \lim_{r\rightarrow 0}\int \xi^j n_j n^i = 3 \langle \xi^j n_j n^i \rangle_{r \rightarrow 0} \label{Zchange}.
\end{equation}
Beginning with the equation of motion in a parity-regular gauge, we may now derive an equation for $Z^i$ in a new gauge,
\begin{align}
\ddot{Z}^i - & \langle F^i \rangle_{r \rightarrow 0} + R_{0j0}^{\ \ \ i} Z^j - M^{-1} S^{kl} R_{kl0i} = \nonumber \\
& \langle ( -\partial_0 \partial_0 \xi_j (\delta^i_{\ j} - 3n^i n_j) + R_{0j0k}x^k \partial^j \xi^i - R_{0i0}^{\ \ \ j}\xi_k (\delta^k_{\ j} - 3 n^k n_j) ) \rangle_{r \rightarrow 0}, \label{eom-noparity}
\end{align}
where equation \eqref{Fchange} has been used.  If the parity condition is satisfied, and $\xi^i = c^i + \Sigma^i(\vec{n})+O(r)$ with $\Sigma^i$ odd-parity, then the right hand side vanishes and the equation of motion retains the original form, depending only on the local spacetime metric (at zeroth and first order in perturbation theory).  However, if the parity condition is not satisfied, then the right hand side does not in general vanish,\footnote{For example, with $\xi^i = B^j n_j n^i$ with $B^i$ constant, the right hand side of \eqref{eom-noparity} evaluates to $(16/15)R_{0i0j}B^j$.} and the equation of motion for $Z^i$ takes a complicated form involving the gauge transformation to some reference gauge.  Another way to see the difficulty is to repeat the calculations of \eqref{dZdd} for a general gauge transformation, giving
\begin{equation}
\delta \ddot{Z}^i = 3 \langle ( \delta F_j + R_{0k0l} x^l \partial_k \xi_j - R_{0j0k}\xi^k ) n^j n^i \rangle_{r\rightarrow 0}.
\end{equation}
Without the parity condition the Riemann terms do not simplify into the geodesic deviation form $R_{0i0j}\delta Z^j$.  It appears that no expression in terms of just $\delta Z^i$ and $\delta h_{\mu \nu}$ is possible, so that the equation for $Z^i$ in parity-irregular gauges must involve a gauge vector explicitly.  In particular, there appears to be no natural separation of the terms contributing to $\ddot{Z}^i$ into ``self'' and other forces in the case of a parity-irregular gauge.

\end{document}